\documentclass[11pt]{article}

\makeatletter
\def\extrarowheight#1{\noalign{\@tempdima\ht\@arstrutbox
\advance\@tempdima#1\ht\@arstrutbox\@tempdima}}
\makeatother

\usepackage{amssymb,amsmath}
\usepackage{cite,soul,marginnote}
\usepackage{graphics}
\usepackage{color}
\usepackage{epstopdf}
\usepackage{epsfig}
\usepackage{times}
\def\RR{\mathbb R}

\newcommand{\norm}[1]{\left\Vert #1 \right \Vert}
\usepackage{mathtools}

\newcommand{\rem}[1]{}
\DeclareMathAlphabet{\mathbi}{OML}{cmm}{b}{it} 
\newcommand{\non}{\nonumber}
\newtheorem{theorem}{Theorem}

\newcommand{\bx}{\mathbi{x}}

\newcommand{\bel}{\begin{equation}\label}
\newcommand{\ee}{\end{equation}}
\newcommand{\beq}{\begin{eqnarray}\label} 
\newcommand{\eeq}{\end{eqnarray}} 
\newcommand{\bc}{\begin{center}} 
\newcommand{\ec}{\end{center}} 
\newcommand{\ben}{\begin{enumerate}}
\newcommand{\een}{\end{enumerate}}
\newcommand{\bit}{\begin{itemize}}
\newcommand{\eit}{\end{itemize}}
\newcommand{\I}{\int_{\mathcal{V}}}

\newcommand{\bdf}{\mathbi{f}}

\newcommand{\bu}{\mbox{\boldmath$u$}}

\newcommand{\bom}{\mbox{\boldmath$\omega$}}

\newcommand\shalf{\ensuremath{{\scriptstyle\frac{1}{2}}}}

\begin{document}

\bc
\textbf{Weak and strong solutions of the $3D$ Navier-Stokes equations and\\
their relation to a chessboard of convergent inverse length scales}
\par\vspace{3mm}
J. D. Gibbon
\par\vspace{3mm}
Department of Mathematics,
\par\vspace{0mm}
Imperial College London, 
\par\vspace{0mm}
London SW7 2AZ, UK.
\ec

\begin{abstract}
Using the scale invariance of the Navier-Stokes equations to define appropriate space-and-time-averaged inverse length scales associated with weak solutions of the $3D$ Navier-Stokes equations, an infinite `chessboard' of estimates for these inverse length scales is displayed in terms of labels $(n,\,m)$ corresponding to $n$ derivatives of the velocity field in $L^{2m}$. The $(1,\,1)$ position corresponds to the inverse Kolmogorov length $Re^{3/4}$. These estimates ultimately converge to a finite limit, $Re^3$, as $n,\,m\to \infty$, although this limit is too large to lie within the physical validity of the equations for realistically large Reynolds numbers. Moreover, all the known time-averaged estimates for weak solutions can be rolled into one single estimate, labelled by $(n,\,m)$. In contrast, those required for strong solutions to exist can be written in another single estimate, also labelled by $(n,\,m)$, the only difference being a factor of 2 in the exponent. This appears to be a generalization of the Prodi-Serrin conditions for $n\geq 1$.
\end{abstract}

\section{\small Introduction}\label{intro}

Numerical simulations of the $3D$ Navier-Stokes equations show that finer and finer vortical structures appear as resolution increases involving  inverse scales much larger than $\lambda_{k}^{-1}$, the inverse Kolmogorov length \cite{DYS2008,IGK2009,Kerr2012}. One widely held belief is that this process is fractal \cite{Mandel1974}\,: that is, the inverse length scales associated with the cascades of vorticity and strain diverge to infinity. In contrast, others believe that viscosity ultimately halts this process \cite{LFR1922,Frisch1995}. L. F. Richardson's neat parody of Jonathan Swift's (de Morgan's version) 
\textit{``Great fleas have little fleas upon their backs to bite 'em, And little fleas have lesser fleas, and so ad infinitum''}, re-expressing it as \textit{``Big whirls have little whirls that feed on their velocity, and little whirls have lesser whirls and so on to viscosity}'', encapsulates both views \cite{LFR1922,Frisch1995,PL2006,RN2013}. A generation ago this question exercised the minds of those who worked at the interface between dynamical systems and fluid turbulence \cite{Frisch1995}. It cannot be wholly settled while the $3D$ Navier-Stokes regularity problem remains open, but it can be attempted for weak solutions, which exist only in the sense that they can be integrated against a test function and are mathematically manifest in terms of estimates of finite time-averages of the velocity field in various function spaces \cite{Leray1934,FGT1981}. 
\par\smallskip
The first task of this paper lies in \S\ref{weaksect}, where it is shown that there exists a bounded, hierarchy of space-and-time-averaged inverse length scales which can be thought of as being associated with a cascade of energy through the scales. The next task is to show that this hierarchy has a finite limit in the sense that the exponent of the Reynolds number ($Re$) in the upper bounds converge. However, this limit is too large to lie within the physical validity of the Navier-Stokes equations at realistic values of $Re$. To follow this process through requires a proper definition of a set of spatially and temporally averaged length scales associated with scaling properties of the Navier-Stokes equations. Ideally this set should be dictated by scale invariance properties, with a free parameter $\lambda$\,: in other words the PDEs themselves should tell us what scales to use.  It turns out that Leray's weak solution formulation, in combination with the natural invariance scaling of the problem, is the natural setting for this problem.  At the first square in a chessboard of scales is the $Re^{3/4}$ Navier-Stokes result of Doering and Foias for the inverse Kolmogorov length $L\lambda_{k}^{-1}$ , which exists as an upper bound for all weak solutions \cite{DF2002}.
\par\smallskip
In \S\ref{wssect} lies the second task of this paper, which is to show that through the scaling invariance results of \S\ref{weaksect}, all the known special cases of $3D$ Navier-Stokes weak solutions \cite{Leray1934,DF2002,Adams1975,Hopf1951,Prodi1959,Serrin1963,Tartar1978,FGT1981,CF1988,DG1995,FMRT2001,ESS2003,RRS2016} can be summarized in one estimate. This is displayed in the first part of Theorem 2, while the second contains the sufficient conditions required to prove the existence of strong solutions. The two estimates are remarkably similar, but with a crucial factor of 2 in the exponent of the latter. This appears to be a generalization of the Prodi-Serrin conditions for $n\geq 1$ \cite{Prodi1959,Serrin1963}. 
\par\smallskip
The Navier-Stokes equations, with $L^2$-bounded body forcing $\bdf(\bx)$, are given by 
\bel{NSE}
(\partial_{t} + \bu\cdot\nabla)\bu = \nu \Delta\bu - \nabla p + \bdf\,,\qquad\qquad \mbox{div}\,\bu = 0\,.
\ee
Weak solutions are available in time-averaged form and it is to these we are restricted if we wish for explicit estimates  \cite{Leray1934,DF2002,Adams1975,Hopf1951,Prodi1959,Serrin1963,Tartar1978,FGT1981,CF1988,DG1995,FMRT2001,ESS2003,RRS2016}. Thus, while we have no control over $\I|\nabla\bu|^{2}dV$ point-wise in time, which is a necessary and sufficient condition for existence and uniqueness, we do have its time average up to time $T$ denoted by
\bel{timavdef}
\left<\cdot\right>_{T} = \frac{1}{T}\int_{0}^{T}\cdot\,dt\,,
\ee
for arbitrarily large $T>0$. This is derived from Leray's energy inequality \cite{Leray1934}. The standard energy dissipation rate \textit{per unit volume} result
\bel{pr1a}
\varepsilon = \nu L^{-3}\left<\I|\nabla\bu|^{2}dV\right>_{T} \leq L^{-4}\nu^{3}GrRe + O\left(T^{-1}\right)\,,
\ee
comes from a direct integration of this inequality. In inequality (\ref{pr1a}) the Reynolds number $Re$, together with the Grashof number $Gr$, are defined by \cite{DF2002}
\bel{Grdef}
Re = \frac{U_{0}L}{\nu}\qquad\qquad Gr = \frac{L^{3}f_{rms}}{\nu^{2}}\,,
\ee
where $U_{0}^{2} = L^{-3}\left<\|\bu\|_{2}^{2}\right>_{T}$ and $f^{2}_{rms} = L^{-3}\|\bdf\|_{2}^{2}$. Doering and Foias \cite{DF2002} have shown that for a forcing concentrated near one wave-number, then $Gr \leq c\,Re^{2}$. With the usual definition 
$\lambda_{k}^{-4} = \varepsilon/\nu^{3}$ they found 
\bel{DF}
L\lambda_{k}^{-1} \leq c\,Re^{3/4}\,.
\ee 

\section{\small Navier-Stokes length scales and weak solutions}\label{weaksect}

\subsection{\small Scale invariance properties}

The Navier-Stokes equations satisfy the scale invariance property
\bel{hscal1}
\bx' = \lambda^{-1}\bx\,;\qquad t' = \Omega t\,;\qquad \bu' = \lambda^{-1}\Omega^{-1}\bu\,;\qquad \nu' = \lambda^{-2}\Omega^{-1}\nu\,,
\ee
where $\lambda$ and $\Omega$ are constant and positive. There are also appropriate scalings on the pressure and forcing.  If $\lambda$ is dimensionless, then the most natural choice for a dimensionless $\Omega$ is $\Omega = \lambda^{-2}$.  This leads to the well-known invariance property
\bel{scin}
\bu = \lambda^{-1}\bu'\left(\bx/\lambda\,,t/\lambda^{2}\right)\,.
\ee
Now consider the  scaling in (\ref{hscal1}) applied to volume integrals of $n$ weak derivatives of the Navier-Stokes velocity field $\bu$ with increasing powers, $2m$ (say), for $1 \leq m \leq \infty$ and $n\geq 0$
\bel{Hnmdef}
H_{n,m} = \I |\nabla^{n}\bu|^{2m}dV \equiv \|\nabla^{n}\bu\|_{L^{2m}}^{2m}\,. 
\ee
Increasing the value of $n$ digs down to smaller scales in the flow, while raising $m$ amplifies the effect of the greater spikes in the data.
\par\smallskip
\renewcommand{\arraystretch}{1.3}
\begin{table}[ht]
\bc
{\small 
\begin{tabular}{||c|c|c|c|c|c|c||}
\hline\hline
$\infty$ & 3 & 3 & 3 & 3 & 3 & 3 \\\hline
$\vdots$ &$\vdots$ &$\vdots$ &$\vdots$ &$\vdots$ &$\vdots$ & $\vdots$\\\hline
4 & $\frac{21}{10}$ & $\frac{51}{20}$ & $\frac{27}{10}$ & $\frac{111}{40}$  & $\ldots$ & 3  \\\hline 
3 & $\frac{15}{8}$ & $\frac{39}{16}$ & $\frac{21}{8}$ & $\frac{87}{32}$ & $\ldots$ & 3 \\\hline 
2 & $\frac{3}{2}$ & $\frac{9}{4}$ & $\frac{5}{2}$ & $\frac{21}{8}$  & $\ldots$ & 3 \\\hline 
1 & $\frac{3}{4}$ & $\frac{15}{8}$ & $\frac{9}{4}$ & $\frac{39}{16}$  & $\ldots$ & 3 \\\hline\hline
$n\uparrow$~$m\to$ & 1 & 2 & 3 & 4 & $\ldots$ & $\infty$\\\hline\hline
\end{tabular}
\caption{\scriptsize A chessboard of length scales\,: entries correspond to the values of the upper bounds on $\left<L\lambda_{n,m}^{-1}\right>_{T}$ 
for integer $(n,m)$ in the range $1\leq n,m\leq 4$. As $n,m\to\infty$ these values converge to the value of 3 the size of which, at realistic Reynolds numbers, is 
likely to be too large to lie within the validity of the Navier-Stokes equations.}\label{tab1}
}
\ec
\end{table}\noindent
The volume integrals defined in (\ref{Hnmdef}) re-scale as
\bel{scal3a}
H_{n,m} = \lambda^{-2m(n-1) +3}\Omega^{2m}H_{n,m}'
\ee
which, with $\Omega = \lambda^{-2}$, gives
\bel{scal3b}
\|\nabla^{n}\bu\|_{L^{2m}} = \lambda^{-1/\alpha_{n,m}}\|\nabla'^{n}\bu'\|_{L^{2m}}'\,,
\ee
where $\alpha_{n,m}$ is defined as 
\bel{alphanmdef}
\alpha_{n,m} = \frac{2m}{2m(n+1)-3}\,. 
\ee
Note that the time integral
\bel{Mudef}
M_{n,m,T}(\bu) = \int_{0}^{T}\|\nabla^{n}\bu\|_{L^{2m}}^{2\alpha_{n,m}}\,dt
\ee
is invariant under the transformation in (\ref{scin}), a fact that will be used later in the proof of Theorem \ref{weakstrong}.

\subsection{\small Definition of a hierarchy of length scales}

To extract an appropriate definition of a set of length scales requires $\lambda$ to be endowed with the dimension of a length\footnote{In this case where 
$\Omega = \nu \lambda^{-2}$ the primed version of the Navier-Stokes equations has unit Reynolds number.} and $\Omega$ the dimension of a frequency. 
The scaling arguments used to derive (\ref{scal3a})  suggest we can define a set of time-dependent length scales $\{\lambda_{n,m}(t)\}$, and a set of 
frequencies $\{\Omega_{n,m}(t)\}$, connected in the following manner
\bel{scal4}
\lambda_{n,m}^{-2m(n-1) + 3}\Omega_{n,m}^{2m} = (L/\lambda_{n,m})^{-3}H_{n,m}\,.
\ee
The dimensionless scaled inverse volume of the domain $(L/\lambda_{n,m})^{-3}$ on the right hand side has been inserted to be sure that the energy 
dissipation rate $\varepsilon$ at the base level $n=m=1$ is properly defined as a \textit{rate per unit volume}, as in the definition of $\varepsilon$  in 
(\ref{pr1a}). So far, the two sets $\{\lambda_{n,m}\}$ and $\{\Omega_{n,m}\}$ have been left independent. To connect them, the most natural 
choice involves $\nu$ and is
\bel{def1}
\Omega_{n,m} = \nu \lambda_{n,m}^{-2} = \varpi_{0} \left(L\lambda_{n,m}^{-1}\right)^{2}\,,
\ee
where $\varpi_{0} = \nu L^{-2}$ is the box frequency. After re-arrangement of (\ref{scal4}), we have the definition
\bel{scal5b}
\left[L\lambda_{n,m}^{-1}(t)\right]^{(n+1)} := \nu^{-1}L^{1/\alpha_{n,m}}\|\nabla^{n}\bu\|_{L^{2m}}\,.
\ee
The proof of the following theorem is given in Appendix \ref{appA} in which $c_{n,m}$ is a generic constant. 
\par\vspace{0mm}\noindent
\begin{theorem}\label{theorem1}
The set of inverse length scales $\lambda_{n,m}^{-1}$ satisfy
\beq{Athm1a}
\left<\left(L\lambda_{n,m}^{-1}\right)^{(n+1)\alpha_{n,m}}\right>_{T} \leq c_{n,m}Re^{3} + O\left(T^{-1}\right)\,,
\eeq
for\,: (i) $n \geq 1$ and $1 \leq m \leq \infty$,~~(ii) $n=0$ and $3 < m \leq \infty$. 
\end{theorem}
\textbf{Remark 1\,:} In fact, because $(n+1)\alpha_{n,m} > 1$, (\ref{Athm1a}) implies the simpler result
\bel{Athm1c}
\left<L\lambda_{n,m}^{-1}\right>_{T} 
\leq c_{n,m}Re^{\frac{3}{(n+1)\alpha_{n,m}}} + O\left(T^{-1}\right)\,.
\ee
Table \ref{tab1} displays the value of the exponent of $Re$ in (\ref{Athm1c}) like bounds on a chessboard for integer values of $(n,\,m)$. 
When $m=n=1$, the conventional Kolmogorov length result is 
\bel{Kol}
\left<L\lambda_{1,1}^{-1}\right>_{T} \leq c\,Re^{3/4} + O\left(T^{-1}\right)\,,
\ee
which is the Doering and Foias result \cite{DF2002}. Thus the first square in the chessboard in Table 
\ref{tab1} is indeed the Kolmogorov estimate $Re^{3/4}$,   
while in the limit $n,\,m \to \infty$
\bel{extra2}
\frac{3[2m(n+1)-3]}{2m(n+1)} \to 3\qquad\mbox{as}\qquad n,\,m \to \infty\,.
\ee
Thus the \textit{mathematical} estimate for these inverse scales does not diverge in the limit\footnote{The constants $c_{n,m}$ 
need more work to determine their exact nature but they converge as $m\to\infty$ but diverge as $n\to\infty$.}. Molecular scales 
lie at about about $0.05\mu$m for air while the corresponding Kolmogorov length is about 1mm. The applicability of the Navier-Stokes 
equations breaks down near molecular scales which the higher entries in Table 1 suggest will be reached even for modest values of $Re$, 
although smaller values of $(n,\,m)$ might be appropriate.
\par\smallskip\noindent
\textbf{Remark 2\,:} An alternative is to express the Theorem in terms of $\Omega_{n,m}$ by defining 
\bel{Dnmdef}
D_{n,m}= \left(\varpi_{0}^{-1}\Omega_{n,m}\right)^{\shalf(n+1)\alpha_{n,m}} = 
L\nu^{-\alpha_{n,m}}\|\nabla^{n}\bu\|_{L^{2m}}^{\alpha_{n,m}}\,,
\ee
in which case 
\bel{Athm1b}
\left<D_{n,m}\right>_{T} \leq c_{n,m}Re^{3} + O\left(T^{-1}\right)\,.
\ee
For the case $n=1$ in (\ref{Athm1b}) we have an estimate for $\left<D_{1,m}\right>_{T}$, which was first found in 
\cite{JDG2011} and investigated numerically in \cite{Donzis2013,Gibbon2014,DG2002}.


\section{\small A contrast between weak and strong solutions}\label{wssect}

Let us now state a Theorem that contrasts what is known for weak solutions compared to what is required for the existence of strong 
solutions. The entries in Table \ref{table2} are many of the well-known special cases already exhibited in the literature for different 
values of $n$ and $m$,\: 
\begin{theorem}\label{weakstrong}
(i) For $n \geq 1$, and $1 \leq m \leq \infty$, and (ii) for $n=0$ and $3 < m \leq \infty$, weak solutions of the $3D$ NSE obey
\bel{w1}
\left<\|\nabla^{n}\bu\|_{L^{2m}}^{\alpha_{n,m}}\right>_{T} \leq c\,L^{-1}\nu^{\alpha_{n,m}}Re^{3} + O\left(T^{-1}\right)\,.
\ee
(iii) For $n \geq 1$ and $1 \leq m \leq \infty$ and (iv) for $n=0$ and $3/2 \leq m \leq \infty$, sufficient conditions for strong solutions of the $3D$ Navier-Stokes equations exist if, for arbitrarily large values of $T$,
\bel{s1}
\left<\|\nabla^{n}\bu\|_{L^{2m}}^{2\alpha_{n,m}}\right>_{T} < \infty\,.
\ee
\end{theorem}
\textbf{Remark 1\,:} In fact, the $3D$ Navier-Stokes equations are regular if (\ref{s1}) holds for a single pair of values $(n,\,m)$ in the advertised ranges. Note the similarity and difference between (\ref{w1}) and (\ref{s1})\,:  (\ref{w1}) is there simply for contrast, while the crucial difference is the factor of 2 in the exponent of (\ref{s1}). The proof is given in Appendix \ref{wsproof}.  Although Theorem \ref{weakstrong} contains no more regularity than any of the conventional results known for many years, as displayed in Table \ref{table2}, it summarizes all of these by rolling them into a single result for weak solutions and then another for prospective strong solutions, with the factor of 2 in the exponent being the main difference. 
\par\smallskip\noindent
\textbf{Remark 2\,:} (\ref{s1}) can also be re-written as a finite bound on the time integral $M_{n,m,T}(\bu)$ already defined in (\ref{Mudef}) andd repeated here
\bel{s2}
M_{n,m,T}(\bu) = \int_{0}^{T}\|\nabla^{n}\bu\|_{L^{2m}}^{2\alpha_{n,m}}\,dt < \infty\,,
\ee
which is invariant under the scaling in (\ref{scin}). This means that what happens at one scale, happens at every scale. The Table displays various examples.
\ben
\item (\ref{s2}) incorporates the Prodi-Serrin conditions at the level of $n=0$ for $m>3$. These are the following \cite{Prodi1959,Serrin1963}\,: for regularity one requires $\bu \in L^{q}\left([0,\,T];\,L^{p}\right)$ such that $2/q + 3/p \leq 1$. This latter expression becomes
\bel{s3}
\frac{2}{2\alpha_{0,m}} + \frac{3}{2m} = \frac{2m-3}{2m} + \frac{3}{2m} = 1\,.
\ee

\item It also incorporates the equivalent of the Beale-Kato-Majda $3D$ Euler criterion at the levels of $n=1$ and $m=\infty$, which is also valid for the Navier-Stokes equations \cite{BKM1984}. At this latter level (\ref{s2}) becomes
\bel{s4}
\int_{0}^{T}\|\nabla\bu\|_{L^{\infty}}dt < \infty\,.
\ee
In turn, if (\ref{s4}) is true, then this implies that $\int_{0}^{T}\|\bom\|_{L^{\infty}}dt < \infty$.

\item  For $n=0$, in the range $1\leq m\leq 3$, the $\alpha_{n,m}$-scaling does not apply for weak solutions, although an estimate has been found there 
by Constantin \cite{Const1991},  where $\alpha_{0,m} = \frac{2m}{2m-3}$ is replaced by $\beta_{m}= \frac{4m}{3(m-1)}$\,: thus $\alpha_{0,m} < 
\beta_{m} < 2\alpha_{0,m}$.
\een
Thus we conclude that $M_{n,m,T}(\bu)$ is a generalization of well-known regularity criteria. 
\par\smallskip
\renewcommand{\arraystretch}{1.3}
\begin{table}[ht]%
\bc
{\scriptsize
\begin{tabular}{||c|c|c|c||}\hline
$\mathbf{n,~m}$ & $\mathbf{\alpha_{n,m} = \frac{2m}{2m(n+1) - 3}}$ & \textbf{Known for weak} & \textbf{Required for strong}\\\hline\hline
$n=0,~m=\infty$ & $\alpha_{0,\infty} = 1$ & $\left<\|\bu\|_{L^{\infty}}\right>_{T} \leq c\,L^{-1}\nu Re^{3}$ \cite{Tartar1978} 
& $\left<\|\bu\|_{L^{\infty}}^{2}\right>_{T} < \infty$\\\hline
$n=0,~m>3$ & $\alpha_{0,m} = \frac{2m}{2m-3}$ & $\left<\|\bu\|_{L^{2m}}^{\alpha_{0,m}}\right>_{T} 
\leq L^{-1}\nu^{\alpha_{0,m}}Re^{3}$ &  $\left<\|\bu\|_{L^{2m}}^{2\alpha_{0,m}}\right>_{T} <\infty$\\\hline
$n=0,~m>3/2$ &$\alpha_{0,m} = \frac{2m}{2m - 3}$ & & Prodi-Serrin $\frac{2}{2\alpha_{0,m}} + \frac{3}{2m} = 1$ \cite{Prodi1959,Serrin1963}\\\hline
$n=0,~m=3/2$ & $\alpha_{1,3/2} = 1$ & &  $\|\bu(\cdot,\,t)\|_{L^{3}} \leq c\,\int_{0}^{t}\|\nabla\bu\|_{L^{3}}^{2}\,d\tau$ \cite{ESS2003}\\\hline
$n=1,~m=\infty$ & $\alpha_{1,\infty} = 1/2$ & $\left<\|\nabla\bu\|_{L^{\infty}}^{1/2}\right>_{T} \leq c\,L^{-1}\nu^{1/2} Re^{3}$ 
\cite{FGT1981}& $\left<\|\nabla\bu\|_{L^{\infty}}\right>_{T} < \infty$ \\\hline
$n=1,~m=1$ & $\alpha_{1,1} = 2$ & $\left<H_{1,1}\right>_{T} < c\,\nu^{2}L^{-1}Re^{3}$ \cite{Leray1934,DF2002} 
& $\left<H_{1,1}^{2}\right>_{T} < \infty$\\\hline
$n=1,~m\geq 1$ & $\alpha_{1,m} = \frac{2m}{4m-3}$ & $\left<D_{1,m}\right>_{T} \leq c\,Re^{3}$\cite{JDG2011} & 
$\left<D_{1,m}^{2}\right>_{T} < \infty$\\\hline 
$n\geq 1,~m=1$ & $\alpha_{n,1} = \frac{2}{2n-1}$ & $\left<H_{n,1}^{\frac{1}{2n-1}}\right>_{T} \leq \nu^{\alpha_{n,1}}L^{-1}Re^{3}$ 
\cite{FGT1981} & $\left<H_{n,1}^{\frac{2}{2n-1}}\right>_{T} < \infty$\\\hline
\hline
\end{tabular}
\caption{\scriptsize A list of (\ref{w1}) and (\ref{s1}) for various specific values of $n,~m$ that correspond to well known results.  The $D_{1,m}$ 
in row 7, column 3 are defined in (\ref{Dnmdef}). See item 3 below equation (\ref{s4}) for Constantin's weak solution estimate \cite{Const1991} 
for the range $1 \leq m \leq 3$.}
\label{table2}
}
\ec
\end{table}


\section{\small Conclusion}\label{con}

The conventional idea of a cascade of energy through the scales has been part of the abiding folk-lore in turbulence, but defining exactly how these scales could be defined and rigorously associated with $3D$ Navier-Stokes solutions has been an open problem until now. It is clear from the definition of the set of  time-averaged inverse length scales $\{\lambda_{n,m}^{-1}\}$, and the results of Theorems \ref{theorem1} and \ref{weakstrong}, that these are closely associated with the full range of weak solutions. Not surprisingly, their definition is also intimately connected with Navier-Stokes invariance properties, although it also raises the question 
from where bounds for strong solutions, i.e. the requirement that $M_{n,m,T}(\bu) < \infty$, could originate? Computing the dynamics at these scales is a challenge, as has been shown by the work in \cite{Donzis2013,Gibbon2014} in which the $D_{1,m}(t)$ were computed as functions of $t$ for $m = 1, ...\,, 9$. How high one can go in $n$ is not clear.
\par\smallskip
While we cannot assume that a full $3D$ Navier-Stokes flow, evolving from arbitrary initial data, is homogeneous and isotropic, nevertheless the results of our analysis touch those of Kolmogorov's 1941 theory in two respects \cite{Frisch1995}.  Firstly, the $Re^{3/4}$-estimate in the first square on the chessboard of the set $\{L\lambda_{n,m}^{-1}\}$ coincides with Kolmogorov length \cite{DF2002}. The other traditional length scale in turbulence is the inverse Taylor micro-scale $L\lambda_{tms}^{-1}\sim Re^{1/2}$\,: here $L$ is the outer scale or box-size of the system in question. This too can be recovered by consulting (\ref{Grdef}) and writing
\bel{lamtms}
\left(L\lambda_{tms}^{-1}\right)^{2} := \frac{\left<\|\bom\|_{2}^{2}\right>_{T}}{\left<\|\bu\|_{2}^{2}\right>_{T}} 
\leq c\,Re + O\left(T^{-1}\right)\,,
\ee
which gives the traditional $Re$ upper bound.  The second touching point is the use of scaling, even though this is significantly different in the two cases. Frisch's book shows how  K41 theory takes as its two free parameters $(\lambda,\,h)$ with $\Omega = \lambda^{h-1}$ \cite{Frisch1995}. The scaling transformation (\ref{hscal1}) then makes $\nu' = \lambda^{-1-h}\nu$ with $\bu' = \lambda^{-h}\bu$, while $\varepsilon' = \lambda^{1-3h}\varepsilon$.  The  requirement that the latter is constant makes $h=1/3$. The Parisi-Frisch multi-fractal model then allows a departure from $h=1/3$ in a probabilistic sense \cite{Frisch1995,PF1985,BenziBif2009}. In contrast, at the heart of the definition of $\lambda_{n,m}$ for the Navier-Stokes equations is the invariance scaling given in (\ref{hscal1}) with $\Omega = \lambda^{-2}$, which reduces the system to one free parameter. It is this scaling that generates the exponent $\alpha_{n,m}$ and it is this exponent that also lies at the heart of both Theorems \ref{theorem1} and \ref{weakstrong}, with the latter expressing estimates for both weak and strong solutions with only a factor of 2 difference between them. A departure from invariance in the form $\Omega = \lambda^{-2 + \delta}$, thus restoring a two-parameter freedom, would be an interesting future topic to discuss.

\par\bigskip\noindent
\textbf{Acknowledgements\,:} My thanks go to Vlad Vicol of Princeton University for suggesting the method of proof of Theorem \ref{weakstrong} and to 
Darryl Holm for discussions.

\appendix
\section{\small Appendix\,: Proof of Theorem \ref{theorem1}}\label{appA}

The following proof is based around the independent result of Foias, Guillop\'e and Temam \cite{FGT1981} on 
the bounded hierarchy of time averages
\bel{chess2}
\left<H_{n,1}^{\frac{1}{2n-1}}\right>_{T} \leq c_{n}L^{-1}\nu^{\frac{2}{2n-1}}Re^{3} + O\left(T^{-1}\right)\,.
\ee
$H_{n,1}$ is defined by (\ref{Hnmdef}) with $m=1$. Given the nature of the $H_{n,m}$ defined in (\ref{Hnmdef}) we are 
dealing with $W^{n,2m}$-spaces with initial data $\bu_{0} \in \dot{W}^{n,2m}$. The aim is to show that 
$\bu \in L^{\alpha_{n,m}}\left([0,\,T],;~\dot{W}^{n,2m}(\RR^{3})\right)$ for $n\geq 1$ with $1 \leq m \leq \infty$ 
and for $n=0$ with $3 < m \leq \infty$.

\subsection{\small The case $n\geq 1$ and $1\leq m \leq \infty$}\label{ngerone}

In terms of time-averages we wish to estimate $\left<\|\nabla^{n}\bu\|_{L^{2m}}^{\alpha_{n,m}}\right>_{T}$ when 
$n\geq 1$.  We first use the Gagliardo-Nirenberg inequality (with $\mathcal{A}\equiv \nabla\bu$) to interpolate between 
$\|\nabla^{n-1}\mathcal{A}\|_{L^{2m}}$ and $\|\nabla^{N}\mathcal{A}\|_{L^{2}}$ \cite{Adams1975}
\bel{pr2}
\|\nabla^{n-1}\mathcal{A}\|_{L^{2m}} \leq C\, \|\nabla^{N}\mathcal{A}\|_{L^{2}}^{a}\|\mathcal{A}\|_{L^{2}}^{1-a}\,,
\ee
where the standard dimensional formula for $a$ is 
\bel{adef2}
a = \frac{m(2n+1)-3}{2mN}.
\ee
We require $(n-1)/N \leq a < 1$ so $N$ must be chosen such that $N > \shalf(2n+1)-3/2m$. The other end of the 
inequality is automatically satisfied for $m \geq 1$. Now we introduce the exponent $\alpha_{n,m}$ and time-average\,:
\beq{pr3}
\left<\|\nabla^{n}\bu\|_{L^{2m}}^{\alpha_{n,m}}\right>_{T} &\leq& c_{N,n,m}
\left<\left(H_{N+1,1}^{a/2}H_{1,1}^{(1-a)/2}\right)^{\alpha_{n,m}}\right>_{T}\non\\
&=& c_{N,n,m}\left<\left\{\left(H_{N+1,1}^{\frac{1}{2N+1}}\right)^{(2N+1)a/2}
H_{1,1}^{(1-a)/2}\right\}^{\alpha_{n,m}}\right>_{T}\non\\
&\leq& c_{N,n,m}\left<H_{N+1,1}^{\frac{1}{2N+1}}\right>_{T}^{\frac{(2N+1)a\alpha_{n,m}}{2}}
\left<H_{1,1}^{\frac{(1-a)\alpha_{n,m}}{2-(2N+1)a\alpha_{n,m}}}\right>_{T}^{\frac{2-(2N+1)a\alpha_{n,m}}{2}}
\eeq
where a H\"older inequality has been used at the last step. We know the first term on the last line of 
the right hand side is bounded using (\ref{chess2}). The last term is bounded only if the exponent of $H_{1,1}$ inside the average is unity 
\bel{pr4}
\frac{(1-a)\alpha_{n,m}}{2-(2N+1)a\alpha_{n,m}} = 1\qquad\Rightarrow \qquad 
\alpha_{n,m} = \frac{2}{2Na+1}\,.
\ee
From (\ref{adef2}) we note that the combination $2Na$ is a function of $m$ only, and gives the correct formula for 
$\alpha_{n,m}$, uniform in $N$. Checking that the coefficients in $L$ and $\nu$ are correct is an exercise in algebra. 
\hfil $\blacksquare$

\subsection{\small The case $n = 0$ and $3 < m \leq \infty$}\label{nezero}

Firstly, we prove a generalized inequality of the type first used by Tartar \cite{Tartar1978} in bounding 
$\left<\|\bu\|_{\infty}\right>_{T}$. We use the Gagliardo-Nirenberg inequality
\bel{GTI1a}
\|\bu\|_{L^{p}} \leq c\,\|\nabla\bu\|^{a}_{L^{2}}\|\bu\|_{L^{2m}}^{1-a}\,,\qquad\qquad a = \frac{3(2m-p)}{p(m-3)}\,,
\ee
where $m > 3$ and $6 < p < 2m$. This ensures that $a < 1$. Next we use another inequality for $N > 3/2$
\bel{GTI2a}
\|\bu\|_{L^{2m}} \leq c\,\|\nabla^{N}\bu\|^{A}_{L^{2}}\|\bu\|_{L^{p}}^{1-A}\,,\qquad\qquad 
A = \frac{3(2m-p)}{m[p(2N-3)+6]}\,.
\ee
Taken together these give\,: for $m > 3$ and $N > \frac{3(m-1)}{2m}$ 
\bel{GTI3a}
\|\bu\|_{L^{2m}} \leq c\,\|\nabla^{N}\bu\|^{B}_{L^{2}}\|\nabla\bu\|_{L^{2}}^{1-B}\,,\qquad\qquad B = \frac{m-3}{2m(N-1)}\,.
\ee
When $m\to\infty$ we recover the $L^{\infty}$-inequality, where $B = \frac{1}{2(N-1)}$.
\par\smallskip
Next we proceed to prove (\ref{Athm1a}) for $n=0$. For an exponent $\alpha_{0,m}$ to be determined we use (\ref{GTI3a}) 
and write
\beq{P1}
\left<\|\bu\|_{L^{2m}}^{\alpha_{0,m}}\right>_{T} &\leq& c_{N,m}\left<H_{N,1}^{B\alpha_{0,m}/2}
H_{1,1}^{(1-B)\alpha_{0,m}/2}\right>_{T}\non\\
&\leq& c_{N,m}\left<\left(H_{N,1}^{\frac{1}{2N-1}}\right)^{B(2N-1)\alpha_{0,m}/2}
H_{1,1}^{(1-B)\alpha_{0,m}/2}\right>_{T}\non\\
&\leq& c_{N,m}\left<H_{N,1}^{\frac{1}{2N-1}}\right>_{T}^{B(2N-1)\alpha_{0,m}/2}\non\\
&\times&\left<H_{1,1}^{\frac{(1-B)\alpha_{0,m}}{2 - B(2N-1)\alpha_{0,m}}}\right>_{T}^{1 - B[2N-1]\alpha_{0,m}/2}\,,
\eeq
To be able to bound the right hand side of (\ref{P1}) from above we use the result of Foias, Guillop\'e and Temam \cite{FGT1981} expressed 
in (\ref{chess2}). In addition, to be able to use the upper bound on $\left<H_{1,1}\right>_{T}$ we set
\bel{P2}
\frac{(1-B)\alpha_{0,m}}{2 - B(2N-1)\alpha_{0,m}} = 1\,,
\ee
which determines $\alpha_{0,m}$. Given that $B = \frac{m-3}{2m(N-1)}$ from (\ref{GTI3a}), we recover
\bel{P3}
\alpha_{0,m} = \frac{2m}{2m-3}\,,\qquad m > 3\,,
\ee 
uniform in $N$, which is the result as advertised. \hfil $\blacksquare$

\section{\small Appendix\,: Proof of Theorem \ref{weakstrong}}\label{wsproof}

The first parts of the theorem $(i)$ and $(ii)$, inequality (\ref{w1}) has already been proved in Theorem \ref{theorem1}. For parts $(iii)$ and $(iv)$, 
involving (\ref{s1}), consider $n$ such that for $n=0$ with $m$ lying in the range $3/2 \leq m\leq \infty$, and $n\geq 1$ 
with $m$  lying the range $1 \leq m \leq \infty$\,; we wish to prove that  strong solutions exist if $\bu \in L^{2\alpha_{n,m}}\left([0,\,T]\,;
~\dot{W}^{n,2m}(\RR^{3})\right)$. Consider (\ref{Mudef}) written down again as
\bel{B1}
M_{n,m,T}(\bu) := \int_{0}^{T} \norm{\nabla^n \bu}_{L^{2m}}^{2 \alpha_{n,m}}\,dt\,.
\ee
Does the assumption $M_{n,m,T}(\bu) < \infty$ imply there is a smooth solution on $[0,\,T]$? 
\par\smallskip\noindent
\textbf{The case $n=0$,~$3/2 < m \leq \infty$\,:} Does the above assumption imply that $\bu \in L^{q}\left([0,\,T];\,L^{p}\right)$? For completeness we 
repeat the argument given in item 1, equation (\ref{s3}). For $n=0$, let $p=2m$ and $q= 2\alpha_{0,m} = 2p/(p-3)$ then $2/q + 3/p = (p-3)/p + 3/p = 1$, 
which is the Prodi-Serrin criterion \cite{Prodi1959,Serrin1963} -- see line 3 of Table \ref{table2} and also (\ref{s2}). The special case $n=0$,~$m=3/2$ is 
dealt with below.  
\par\smallskip\noindent
\textbf{The case $n\geq 1$,~$1 \leq m \leq \infty$\,:} As already pointed out, a simple time integration of  equation (\ref{scal3b}) verifies that $M_{n,m,T}(\bu)$ is  \textit{scaling invariant} for any $n\geq 0$\,; that is, if $\bu(\bx,t)$ is a solution of 3D Navier-Stokes on $\RR^3 \times [0,\,T]$, then under $\bu_{\lambda}(\bx,t) =\lambda^{-1} \bu\left(\frac{\bx}{\lambda},\frac{t}{\lambda^2}\right)$\, we have $M_{n,m,T}(\bu) = M_{n,m,\lambda^2 T}(\bu_{\lambda})$. Thus, requiring the boundedness of the quantity $M_{n,m,T}$ from \eqref{B1} appears to be the  generalization of the Prodi-Serrin criterion for $n\geq 1$. The following is an adaptation of the standard proof of this criterion to the case $n\geq 1$. 
\par\smallskip
For initial data $\bu_{0} \in \dot{W}^{n,2m}$, the norm which is integrated in time in \eqref{B1}, it is not difficult to prove a local existence 
in time result, with the time of existence depending only on the norm of the initial data. Then the proof is by contradiction using the standard maximal 
in time $T_{*} = T_{*}\left(\norm{\bu_{0}}_{\dot{W}^{n,2m}}\right) >0$ argument. Assume that a weak solution $\bu$ of 3D Navier-Stokes obeys $M_{n,m,T}(\bu) < \infty$, but that it blows up at time $T$.  In particular, this means that for \textit{any} increasing sequence of times $t_{n} \nearrow T$ we must have
\bel{B2}
\lim_{n\to \infty} \norm{\bu(\cdot,\,t_n)}_{\dot{W}^{n,2m}} = \infty.
\ee
Otherwise, by the local existence theorem, one can extend the solution past time the putative blowup time $T$. Using this sequence of times we renormalize the solution $\bu$ according to the sequence of times $t_n$ as 
\bel{B3}
\bu_{n}(\bx,t) = \frac{1}{\Lambda_n} \bu\left(\frac{\bx}{\Lambda_n},\,t_{n} + \frac{t}{\Lambda_{n}^{2}}\right)
\ee
where $\Lambda_n>0$ is defined\footnote{We use capital $\Lambda_{n}$ to avoid confusion with small $\lambda_{n,m}$.} such that 
\bel{B4}
\norm{\bu_{n}(\bx,\,0)}_{\dot{W}^{n,2m}} = 1. 
\ee
In particular, we have that $\Lambda_{n} \to \infty$ as $n\to \infty$, because (\ref{scal3b}) and (\ref{B4}) together show that
\bel{B5}
\Lambda_{n} = \norm{\bu(\cdot,\,t_{n})}_{\dot{W}^{n,2m}}^{\alpha_{n,m}}\,.
\ee
Note that the functions $\bu_{n}$ also solve the $3D$ Navier-Stokes equations, but since the solution $\bu$ lives only up to time $T$, we know that the 
solutions $\bu_{n}$ live up to time $(T-t_{n}) \Lambda_{n}^{2}$. Thus, $\bu_{n}$ solves the $3D$ Navier-Stokes equations on $\RR^{3} 
\times [0,\,(T-t_{n}) \Lambda_{n}^{2})$. However, by (\ref{B4}) and the local existence result, we know that $\bu_{n}$ does not 
blow up before the local existence time $T_{*}(1)>0$. Therefore, we must have
\bel{B6}
(T-t_{n}) \Lambda_{n}^{2} \geq T_{*}(1)\,.
\ee
By (\ref{B5}), the above condition implies that 
\bel{B7}
\norm{\bu(\cdot,\,t_{n})}_{\dot{W}^{n,2m}}^{2\alpha_{n,m}}  \geq \frac{T_*(1)}{T-t_n}\,.
\ee
However, the sequence $t_{n}\nearrow T$ was arbitrary, and thus (\ref{B7}) implies a minimal blowup rate
\bel{B8}
\norm{\bu(\cdot,\,t)}_{\dot{W}^{n,2m}}^{2\alpha_{n,m}}  \geq \frac{T_*(1)}{T-t}\,.
\ee
The contradiction is now immediate\,: we have made the assumption $M_{n,m,T}(\bu) < \infty$, and thus the left side of (\ref{B8}) 
is integrable on $[0,\,T)$. On the other hand, the right side of (\ref{B8}) is not integrable on $[0,\,T)$, which is the desired contradiction. 
\par\smallskip\noindent
\textbf{The special case $n=0$,~$m=3/2$\,:} The $\|\bu(\cdot,\,t)\|_{L^3}$ result, excluded by the Prodi-Serrin conditions, but 
proved by Escauriaza, Seregin, and Sver\'ak \cite{ESS2003}, (see row 4 of Table \ref{table2}) can be shown to be controlled by (\ref{s1}) 
when $n=1$ and $m=3/2$. To prove this we ignore the Laplacian term and write
\bel{u3a}
\frac{d~}{dt}\|\bu\|_{L^{2m}} \leq \|\nabla p\|_{L^{2m}}\,.
\ee
We use the Sobolev embedding
\bel{sob1}
\|A\|_{L^{q}} \leq c\|\nabla A\|_{L^{p}}\qquad\qquad\frac{1}{q} = \frac{1}{p} - \frac{1}{3}
\ee
in three dimensions. This fits nicely for $p=3/2$ and $q=3$. In other words $\|A\|_{L^{3}} \leq c\|\nabla A\|_{L^{3/2}}$. 
Thus, with $m=3/2$, we can write
\bel{u3b}
\|\nabla p\|_{L^{3}} \leq \|\nabla^{2} p\|_{L^{3/2}} \leq c\, \|\nabla^{2}\Delta^{-1}u_{i,j}u_{j,i}\|_{L^{3/2}}
\leq c\, \|\nabla\bu\|_{L^{3}}^{2}\,, 
\ee
having used a Riesz transform. Thus
\bel{u3c}
\|\bu(\cdot,\,t)\|_{L^{3}} \leq c\,\int_{0}^{t}\|\nabla\bu\|_{L^{3}}^{2}\,d\tau\,.
\ee
Finally $\alpha_{1,3/2} = \frac{3}{6-3} = 1$, and so $\|\bu(\cdot,\,t)\|_{L^{3}}$ is controlled by (\ref{s1})
when $n=1,~m=3/2$. \hfil $\blacksquare$


\end{document}